# A Multi-Agent based Approach for Simulating the Impact of Human Behaviours on Air Pollution

Sabri Ghazi
Laboratoire de gestion électronique du document, Computer Science department, University Badji Mokhtar, PO-Box 12, 23000,Annaba, Algeria
E-mail: sabri.ghazi@univ-annaba.dz

Julie Dugdale
University Grenoble Alps. LIG. France.
Email: julie.dugdale@imag.fr

Tarek Khadir
Laboratoire de gestion électronique du document, Computer Science department, University Badji Mokhtar, PO-Box 12, 23000,Annaba, Algeria
E-mail: khadir@labged.net

*This paper presents a Multi-Agent System (MAS) approach for designing an air pollution simulator. The aim is to simulate the concentration of air pollutants emitted from sources (e.g. factories) and to investigate the emergence of cooperation between the emission source managers and the impact this has on air quality. The emission sources are controlled by agents. The agents try to achieve their goals (i.e. increase production, which has the side effect of raising air pollution) and also cooperate with others agents by altering their emission rate according to the air quality. The agents play an adapted version of the evolutionary N-Person Prisoners' Dilemma game in a non-deterministic environment; they have two decisions: decrease or increase the emission. The rewards/penalties are influenced by the pollutant concentration which is, in turn, determined using climatic parameters. In order to give predictions about the concentration of pollutants: Particulates Matter (PM10), Sulphur Oxide and Dioxide ($SO_x$), Nitrogen Oxides ($NO_x$) and Ozone: ($O_3$), a two stage prediction method is used, a GPD (Gaussian Plume Dispersion) model and an ANN (Artificial Neural Network) prediction model. The prediction is calculated using the dispersal information and real data about climatic parameters (wind speed, humidity, temperature and rainfall). Every agent cooperates with its neighbours that emit the same pollutant, and it learns how to adapt its strategy to gain more reward. When the pollution level exceeds the maximum allowed level, agents are penalised according to their participation. The system has been tested using real data from the region of Annaba (North-East Algeria). It helped to investigate how the regulations enhance the cooperation and may help controlling the air quality. The designed system helps the environmental agencies to assess their air pollution controlling policies.*

## 1 Introduction

The question about how humans should moderate their exploitation of environmental resources has occupied researchers for decades [1]. Promoting social and economic growth without affecting the environmental equilibrium is important for maintaining sustainable development. This paper addresses the relation between human behaviours and their impact on air quality in socio-environmental systems. Air pollution is a major concern in many cities in the world, especially in developing countries. It has a direct influence on our health and quality of life [2]. The degradation in air quality should be estimated before the establishment or the expansion of urban or industrial activities. Air pollution simulation and decision support tools can help decision-makers to establish policies for environmental management and to predict the impact of their decisions on the environment and ecosystem. Many modelling approaches have been proposed to study air pollution. Most of them, ([3], [4], [5]) to cite a few, are mainly focused on the physical and chemical aspects of air pollution; the concentration and dispersal of pollutant. These models do not take into consideration human-decision factors. Air pollution is by nature distributed and includes the interaction of individuals involved in the exploitation of a dynamic ecological resource which is the air. The anthropogenic activities (road traffic, industrial and agricultural activities) are among the major sources of air pollution. All of these activities are controlled by humans; therefore, including the human decision factors in the modelling of air pollution is essential.

MAS (Multi-Agent System) based models are an appropriate method for modelling socio-environmental issues [6]. They allow us to model the behaviours of human actors sharing the exploitation of environmental resources. [7] presents a review of recent MAS models used to investigate socio-environmental problems. The models are classified according to: the decision making mechanism, the use or not of real data, the objective of the simulation, and the space and time representation. [8], [9] used a MAS approach to investigate the air pollution emission resulting from road activities; they used a traffic flow simulation and linked it to emission calculation. [10] used the same approach to study the effect of transport regulation on air pollution emission. [11] present a MAS designed for monitoring air quality in Athens, Greece. The MAS is a set of agents that control a network of sensors installed in an urban region. They verify and collect the data measured by sensors. [12] present a MAS to find the dispersion of air pollution in urban region. The pollution sources (polluters) are represented by homogeneous agents that emit pollution in their respective areas. Each agent pollutes with an emission rate. As the simulation runs, clusters are formed with different values of pollution concentration. At the end, a single cluster is formed, thus, the dispersion of pollution is estimated.

The managers of emission sources share the exploitation of the air by emitting pollutants. We aim to simulate their different personalities (e.g. eco-friendly, selfish) and investigate the relationship between the emergence of cooperation and its impact on air quality. The main questions addressed in this paper are: How do emission source controllers cooperate, are they able to achieve their goals while preserving a reasonable air quality? What regulatory rules should be adopted to enhance the cooperation and sustain the air quality? To investigate these questions we have designed a MAS simulation tool that helps to investigate the emergence of cooperation and its effects on air quality. The proposed simulator models the population of emission source controllers as a network of agents playing an Evolutionary NPPD (N-Person Prisoners' Dilemma) game. Evolutionary NPPD has been widely used for studying the emergence of cooperative behaviour among a population of selfish agents, how agents exhibit altruistic behaviours and under which condition cooperation will be sustained.

NPPD is a mathematical model, which models the conflict between players sharing the use of a common resource. Initially, it was formulated for two players where each one has to take two possible actions (defect or cooperate), and then receive a payoff according to their joint actions. A version for N-Persons has been proposed [13] where the payoff is calculated according to the number of agents choosing to cooperate; the payoff function is given in (1)

$$u(ncp) = \begin{cases} \frac{b*ncp}{N} - c & if \ s = 0 \\ \frac{b*ncp}{N} & if \ s = 1 \end{cases} \quad (1)$$

With $b > c > 0 \ and \ \ c > \frac{b}{N}$, $s$ is the action taken by the agent $s \ \epsilon\{0,1\}$ (where 0 means cooperate and 1 means defect), $ncp$ is the number of players who chose to cooperate, $N$ is the size of the player population and $b$ is the defection temptation, the constant $c$ is used to ensure that the cooperation reward is less than the defection reward.

[14], [15], studied the emergence of cooperation in a NPPD. The authors used different agent personalities and neighbourhoods in order to investigate their impact on the evolution of the game outcome. The experiment used different agent types with different initial co-operators ratio; this showed that for the case where all agents are Pavlovians (repeating actions that give them more satisfaction), the aggregate outcome of the game can be predicted for any number of agents and any payoff function. The choice of the agent's neighbours also has a big influence on the game equilibrium. [16] investigated the effect of social welfare preference on the emergence of cooperation among agents placed on a BA [17] network. The authors proposed a model where some of the agents also take into consideration social welfare and not only their payoff received from the game. Agents do not only care about their own payoff, but also the payoff of their neighbours.

[18] describes the use of a NPPD game to investigate the cooperation in a socio geographic community. The use of NPDD for environmental modelling has proved to be suitable since the exploitation of a shared ecological resource can be formulated as a tragedy of the common [19]. Each actor tends to maximise its profits by exploiting a shared ecological resource. Thus, a tragedy of the common arises. [20] uses a PD model to review Porters' hypothesis, which studies the relationship between productivity and eco-friendly technologies. The work models how strict environmental regulations can enhance innovation for a less polluting technology. Firms have two actions which are: to invest in a new less polluting process or to continue using the old one and be penalised according to the governmental regulations. [21] used a version of NPPD to investigate the cooperation in international environmental negotiation to reduce CO2 emissions. [22] presents an evolutionary game theory approach to study the influence of the ecological dynamic and payoff structures over the emergence of cooperative behaviour between landowners. The landowners are modelled as selfish agents aiming to maximize their profit by managing the number of deer on their lands. The main novelty of our approach is the inclusion of human decisions as a key element for simulating the air pollution evolution. We model the managers of the emission sources of pollutants as autonomous agents. These agents aim to maximise their own profit and we investigate this effect on air quality. The designed system helps investigating the efficiency of the regulatory rules used by air pollution controlling agencies for maintaining the air quality. This is very

important because it helps the environmental agencies to assess their air pollution controlling policies.

The paper is organised as follows: The methodology is presented in section (2) that describes a MAS approach for designing an air pollution agent based simulator. Subsection (2.1) presents the representation of space and time. Subsection (2.2) describes the dispersion model and the Artificial Neural Network (ANN) prediction model.

Many conceptualizations have been proposed to represent a socio-environmental system [22]. Generally, a socio-environmental simulation system can be represented as an interconnection of three components (or subsystems); each one is represented by a set of variables (attributes) forming its state at time $t$. The ecological component models the biotic (living) and abiotic (non-living) parts. The economic component represents the economic view point and groups the economic variables. The social component represents the human social networks such as decision-makers, firms, government agencies and consumers. A change in the state variable of each component affects other systems' state variables. For example, the increase in demand for a certain kind of fish, leads fishermen to intensify their exploitation; this in turn results in changes to the biodiversity. We present a generic formalization of a socio-environmental system. A coupled social and environmental system can be expressed as a set of economic, social and ecologic state variables. The state of the system at time step $t$ can be formulated as (2).

$$ES_t = < Ec_t, Sc_t, Envc_t > \text{ (2)}$$

Where $Ec$, $Sc$ and $Envc$ represent, the sets of economic, social and environmental state variables, respectively:

$$Ec_t = < Ec_{1,t}, \dots, Ec_{l,t} >, Sc_t = < Sc_{1,t}, \dots, Sc_{m,t} >$$
$$Envc_t = < Envc_{1,t}, \dots, Envc_{n,t} > \text{(3)}$$

In our case, the environment state variables at time step $t$ are: $Envc_t = < c_{0,t}, \dots c_{n,t}, WS_t, Hu_t, T_t, RF_t >$ (4)

$c_{i,t}$ is the concentration of the pollutant $i$, $WS$: wind speed, $T$: temperature, $Hu$: humidity and $RF$ represents the rainfall, at time $t$. Assuming that the source of pollution at time $t$ is modelled as:

$$S_t = < er_t, tc, X, Y, Z > \text{   (5)}$$

The source produces the pollutant $tc$ with the rate $er$ at the geo-position $(X,Y,Z)$. Sources are controlled by agents. Every agent has to make a decision on which action to choose among all possible actions according to the state of the environment $ES$ and its internal state at $t$. Let $A$ be the set of actions $A = \{a_1, a_2, \dots, a_z\}$, the result of an action is the change in the emission rate of the pollutant from the controlled source. We can define this as a function that takes the agent's action and as a result gives the new emission rate (6).

$$F: A \rightarrow \mathbb{R} \quad \text{(6)}.$$

Let $\pi_t$ be the action vector done by $N$ agents at time $t$:
$$\pi_t = < A_{0t}, \dots A_{nt} > \quad \text{(7)}$$

Let $Q$ be the set of possible air quality index values: $Q = \{very\_bad, bad, average, good, very\_good\}$, each of these indexes has its numerical equivalent in terms of pollutant concentration, as shown in table 1.

The agent decision-making mechanism is given in (2.3). A test scenario is presented in section (3). Results are detailed and discussed in section (4). The paper ends with conclusions and the possible further directions of our work.

## 2   Model Approach and architecture

| SOx µg | NOx µg | O₃ µg | PM10 µg | Indices | Category |
|--------|--------|-------|---------|---------|----------|
| 0 – 30 | 0-45 | 0-45 | 0-20 | 1 | Very Good |
| 30-60 | 45-80 | 45-80 | 20-40 | 2 | Good |
| 60-125 | 80-200 | 80-150 | 40-100 | 3 | Average |
| 125-250 | 200-400 | 150-270 | 100-200 | 4 | Bad |
| >250 | >400 | >270 | >200 | 5 | Very Bad |

Table 1: Air pollution quality

The air quality can be modelled as a graph with $T$ as transition function:

$$T(\pi_t, ES_t, current\_q) \rightarrow new\_q,$$
$$current\_q \text{ and } new\_q \in Q \text{ (8)}$$

$T$ takes as arguments the state of the system $ES_t$ and the set of actions done by $N$ agents and accordingly it moves the system from the current state ($current\_q$) to a new state ($new\_q$). Under some conditions $current\_q$ may be equal to $new\_q$, which means that actions of the agents do not change the air quality under some climatic conditions.

Our simulation approach can be schematised as shown in figure 1. Agents' actions affect the emission rate of the sources. Then the dispersion algorithm is used to compute the dispersion, the aggregated value of pollutant concentration is used with climatic parameters to forecast the next 2 hours air pollution concentration and air quality. According to these forecasts, agents are rewarded or penalised. Agents then adapt their strategies to earn more reward and reduce penalties.

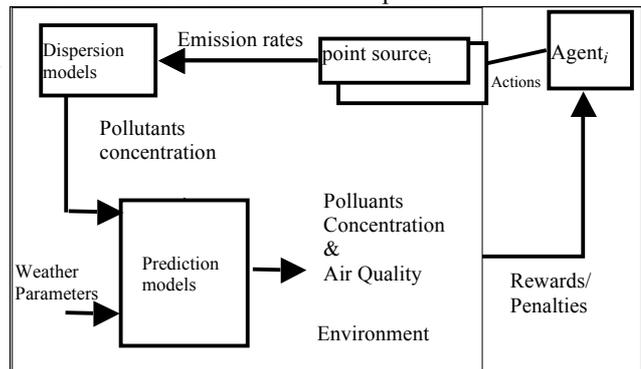

Figure 1: The simulation process, using the dispersion model and the prediction model.

### 2.1   The spatial and temporal scale of the simulation model:

The simulation uses a discrete representation of time where each simulation step represents by default 6 hours of real time. The simulation's duration is defined in the interface and depends on the objective of the simulation (short or long term prediction).

Our model is based on the hypothesis that the action of the emission controllers (reducing or increasing emissions) has an impact within $k$ time-steps. $k$ is a parameter whose value is provided by the user according to the scenario and available data.

Since the simulation step $k$ can be changed, we can represent a long term simulation horizon by giving $k$ a higher value. So for example we can represent 1 step as 24 hours meaning that industrial polluters can take several days to adjust their production volumes. Setting $k$ even higher, such as 2 weeks or 1 month would require data, which is not available, to see the evolution of the *AQ*.

The environment is modelled as a set of *3D boxes*, each one represents one km$^3$ (see figure 3). It can be represented as: $BX = \{bx_0, ..., bx_m\}$, every box is localised in the geo-position point $gp(x,y,z)$ and has attributes representing the concentration of air pollutants $(cp_0,...,cp_v)$ and air quality, These attributes are used when agents are penalized according to the pollution level in the box where they are situated. In this case the position of the emission source in a box is relevant. Sources located in the same box are considered to be neighbours.

## 2.2 Dispersion and prediction models

The dispersion model helps to measure how the pollutant will spread in the air. It is calculated according to the distance from the point source, the wind speed and direction. We used a GPD (Gaussian Plum Dispersion model), which is frequently used in atmospheric dispersion [24]. The dispersion model is run in a steady way, which means that no parameter (wind speed, emission rate and wind direction) is changed during the simulation step. This provides a series of snapshots of the situation at each step. These snapshots are then fed into the ANN model to obtain a prediction about the concentration. Since we cannot combine the two models in a continuous way our solution of taking a series of snapshots and feeding it to the ANN mimics a continuous process. The GPD simulates the dispersal from a point source emission according to the emission rate (9).

$$C(x,y,z,H) = \frac{er_{i,t}*D}{2\pi U_t\,\sigma_y\sigma_z} * e^{-\frac{y^2}{2*\sigma_y^2}} * \left[ (e^{-(\frac{(z-H)^2}{2\sigma_z^2})}) + (e^{-(\frac{(z+H)^2}{2\sigma_z^2})}) \right] \quad (9)$$

This means that the concentration of the pollutant at point $(x,y,z)$ is calculated according to :

$er_{i,t}$: the emission rate in kilograms per hour of the source $i$ in time step $t$.

$U_i$: the wind speed in metres per second at time step $t$, $\sigma_y\sigma_z$: the standard deviation of the concentration distributions in a crosswind in a vertical direction, these two parameters are chosen according to the stability class 'C' in the Guifford-Pasquill scale [25], and $H$ is the height of the source from the ground. The decay term $D$ is given in [26] and computed according to (10).

$$D = \begin{cases} e^{\left(1/(R*\frac{x}{U})\right)}, if \ R > 0 \\ 1, \ if \ R = 0 \end{cases}, \quad (10)$$

Where $x$ is the downwind distance, $u$ is the wind speed and $R$ is the decay coefficient. The values for $R$ are adopted from [27] for NO$_x$ (0.45 h$^{-1}$) and SO$_x$ (0.31 h$^{-1}$), PM10 is not considered ($R=0$ and $D=1$).

For simplification and due to lack of wind direction data, we assume that the wind direction does not change during the simulation step. The resulting pollution level from each source is aggregated and the average of each box is computed. Then the dispersion value of the pollutant is passed to an ANN prediction model as described in [28]. The ANN prediction models are designed to give a forecast of the five air pollutants and the air quality. This includes an uncertainty aspect caused by the weather conditions [29]. The ANN predictor uses the aggregated air pollution concentration value given by the dispersion model of each source and the four climatic parameters: wind speed, humidity, temperature and rain fall. These parameters greatly influence the pollutant concentration in the air [30].

$O_3$ is a secondary pollutant, which means that it is not emitted by sources, but results from the photochemical interaction between SO$_x$, CO$_x$, and organic components. Therefore, we used SO$_x$ and CO$_x$ dispersion information to predict the $O_3$ concentration.

For each pollutant a RBF (Radial Basis Function) network is designed and trained. The RBF is composed of three layers, the first layer is connected to the input of the network and its output is connected to the hidden layer, the neurones in the hidden layer have the RBF as the activation function. The outputs of the hidden layer are linearly combined to obtain the output of the network. Using a training data set, the objective is to find the optimal combination between the number of neurons in the hidden layer and the weight of each input. By increasing the number of the neurons in the hidden layer the algorithm [31] gives the optimal topology of the network. This is why many topologies are tested and only the best of them are taken. During the training step, each network receives as input a vector of the climatic condition parameters and the concentration of the pollutant at time $t$. Each network generates the desired output that is the value of pollutant concentration at time $t+PredictionHorizon$. The forecast given by each network is passed as input to predict the air quality index using a MLP (Multi-Layered Perceptron). The MLP network is trained using the local air quality standards as shown in table 1. Air quality predictions for the different pollutants are obtained on a $t+12$ hours basis and give the most probable air quality category. The MLP model receives the predicted values of the five pollutants, CO$_x$, NO$_x$, $O_3$, PM10, and SO$_x$, and predicts the index values for air quality ranging from 1 to 5, with 1 being very good and 5 being very bad. To train the MLP network we used a Levenberg-Marquardt back-propagation algorithm. The MLP network final topology, obtained after several trials, is: 5 neurons in the first hidden layer, 10 in the second and finally a linear neuron for the output layer.

The accuracy of the ANN models are given in table 2, and is calculated using one year's worth of data according to RMSE (Rooted Mean Squared Error) formulated in (11):

$$RMSE = \sqrt{\frac{\sum_{i=1}^{lng}(P_i - R_i)^2}{lng}} \qquad (11).$$

Where $lng$ is the length of the vectors, $P$ and $R$ are the predicted and measured values, respectively. The performances are computed using a validation data set that was not used in the training of the ANN models.

| Model | Topology [# of input neuron -# of hidden neuron] | Validation error ( RMSE) |
|---|---|---|
| PM10 | [10-320] | 16.1945 µg/m$^3$ |
| SO$_X$ | [10-90] | 3.1618 µg/m$^3$ |
| NO$_X$ | [10- 105] | 9.7277 µg/m$^3$ |
| CO$_X$ | [10-45] | 0.1220 µg/m$^3$ |
| O$_3$ | [10-180] | 39.8238 µg/m$^3$ |

Table 2: Validation error of the ANN prediction models using the validation data set.

## 2.3 Decision-making mechanism

Based on its internal state and the state of the environment, an agent has to choose an action to perform among all possible actions in order to reach its goals. This process is called decision-making. [32] presents a review of methods used for modelling decision-making in a coupled environmental and social system. Our system supports two cooperation strategies (centralized and evolutionary game) each one defines a decision-making mechanism. The centralized strategy (CS) is based on defining a central agent that represents the air pollution controlling agency. The central agent takes decisions according to the current air pollution level. The second strategy is based on an evolutionary game, where agents are rewarded and penalized according to the pollution levels; they make decisions according to their rewards. In our system, the cooperation strategy is defined within the simulation parameters.

### 2.3.1 Centralized Strategy (CS):

The task of maintaining the air quality is assigned to an agent, which represents the air pollution control agency. It uses the GPD and ANN models to predict the air quality and pollutant levels, and according to the predictions it sends a reduce emission message to the emission agents. Then it will check the air quality. It will continue doing this until the air quality is improved to reach the air quality index goal. The central agent has absolute authority and its orders are executed by the emission source controllers. Agents communicate their emission rate at each simulation step to the central agent. This strategy is based on the communication between agents. We assume that agents are rational and have an environmental-responsible personality; this means they favour air quality improvement over their own interests and communicate their exact emission rate to the central agent.

### 2.3.2 Evolutionary Game Cooperating Strategy:

In the EG strategy, every agent has its own goals (earning more rewards and keeping its emission rate high) and shares a global goal of maintaining air quality with other agents. The appreciation function defined as: $app: Q \rightarrow R$, allows comparing the air quality state at each step of the simulation. A global goal GG can be defined as (12). This means finding a set of actions $\pi_t$ to be performed by agents at time $t$, which allows the system to move to a new state of air quality $q_{t+1}$ that is better than the current state.

$$GG_t = \{T(\pi_t, q_t, q_{t+1}), app(q_{t+1}) > app(q_t)\} \quad (12)$$

An agent participates with other agents in the NPPD game, its own goal is to maximise its reward earned from the game. We adopted the approach of [33], where agents keep traces of their $L$ previous steps (actions, rewards and its neighbours' rewards). To update the probabilities to increase or decrease the emission, we used [18] method. At each time step t the agent computes its weighted payoff, according to (13), and tries to maximise it (as its utility function) by taking it into consideration when computing its probability to increase or decrease its emission rate, respectively according to (14) and (15).

$$WP_i(t) = \sum_{i=1}^{L} w_i * M_i(t) \qquad (13)$$

Where: $w_i$ is the weighting parameter where $\sum_{nbr=1}^{L-1} w_{nbr} = 1$ and $\forall i,j \ (i < j \rightarrow w_i > w_j)$, $M_i(t)$ is the payoff for the agent $i$ for the time step $t$.

$$\begin{cases} Pc_i(t+1) = Pc_i(t) + (1 - Pc_i(t) * \alpha_i(t) \ , if \ S_i = 0 \ and \ WP_i(t) > 0 \\ Pc_i(t+1) = (1 - \alpha_i(t)\ ) * Pc_i(t), \qquad if \ S_i = 0 \ and \ WP_i(t) \le 0 \end{cases}$$
$$(14)$$
$$\begin{cases} Pd_i(t+1) = Pd_i(t) + (1 - Pd_i(t) * \alpha_i(t) \ , if \ S_i = 1 \ and \ WP_i(t) > 0 \\ Pd_i(t+1) = (1 - \alpha_i(t)\ ) * Pd_i(t), \qquad if \ S_i = 1 \ and \ WP_i(t) \le 0 \end{cases}$$
$$(15)$$

Where: $Pc_i$ and $Pd_i$ are respectively the probability to decrease ($s=0$) and increase ($s=1$) emissions for agent $i$, $\alpha_i(t)$ is the learning rate of agent $i$ at time step $t$, $s$ is the strategy played at time $t$. The learning rate is updated according to (17):

$$D_i = \sum_{j=1}^{L-1} \begin{cases} 0 \ if \ X_{i,j} = X_{i,j+1} \\ 1 \ if \ X_{i,j} \ne X_{i,j+1} \end{cases} \qquad (16)$$

$$\begin{cases} \alpha_i(t+1) = \alpha_i(t) + 0.015 \qquad if \ D_i = 0 \\ \alpha_i(t+1) = \alpha_i(t) + 0.010 \ if \ D_i > L-1 \\ \alpha_i(t+1) = \alpha_i(t) - 0.010 \ if \ D_i \le L-1 \end{cases} \quad (17)$$

Where $D_i$ is the $i$-th agent actions homogeneity indicator, at time step $t$, $X_{i,j}$ is $j$-th action of the agent $i$. $D_i$ is used to compare the last $L$ actions of the agent. This is used to keep the agent from changing its actions. Agents are influenced by their neighbours, at each time; the average reward of the neighbours is calculated according to (18).

$$nP_i(t) = (\sum_{j=1}^{numberOfneighbours_i} M_j(t)) / numberOfneighbours_i$$
$$(18)$$

Where $M_j(t)$ is the payoff of the neighbour $j$ and $numberOfneighbours_i$ is the number of neighbours for the $i$-th agent. We keep a trace of the $nP$ of the L previous simulation steps and we compute their average in $avgnP$. The agent then uses the probabilities $Pc$, $Pd$ and the average reward of its neighbours to choose an action according to (19):

$$\begin{cases} if\ S_i(t) = 0, S_i(t+1) = \begin{cases} 1, if\ WP_i < avgNP_i\ and\ Pd_i(t+1) > Pc_i(t+1), \\ 0, else \end{cases} \\ if\ S_i(t) = 1, S_i(t+1) = \begin{cases} 0, if\ WP_i < avgNP_i\ and\ Pc_i(t+1) > Pd_i(t+1), \\ 1, else \end{cases} \end{cases}$$
(19)

At each simulation step, every agent gets a reward or penalty according to its actions and according to the pollution level. We have adopted the payoff curve (1) with $b=2$ and $c=-0.5$, but in the general case these parameters can be defined by the user. When the pollution level is higher than the maximum allowed value, the participation of the agent to the current level of the pollution $\sigma_i(t)$ is computed according to (20).

$$\sigma_i(t) = \frac{ER_i(t)}{PL_e(t) - PL_{max}}, PL_e(t) > PL_{max}$$
(20)

Where, $ER_i(t)$ is the emission rate of the $i$-th agent at time $t$, $PL_e(t)$ is the pollution level of the pollutant $e$ at time $t$ and $PL_{max}$ is the maximum allowed value for the pollutant level according to the regulation and local standards. The penalty for agent $i$ at time step $t$ is calculated according to (21):

$$EcoFactor_i(t) = EcoFactor_i(t-1) + (1 - \frac{1}{e^{\sigma_i(t)}})\ (21)$$

Two penalising strategies were used; the first uses (21) and is a cumulative penalty. This means that the penalties from each step are kept and the agent is penalised as long as it continues to increase its emission. The second penalising method is not cumulative, and agents are penalised just according to the current simulation step.

The reward of agent $i$, at the current time step $t$ is computed according to (22), we compute the number of agents who choose to decrease their emission denoted $ncp$, after that we compute $u$ use as defined in equation (1).

$$M_i(t) = \begin{cases} u(ncp) & if\ s_i = 0 \\ u(ncp) - EcoFactor_i(t) & if\ s_i = 1 \end{cases}\ (22)$$

# 3    Simulation scenarios using data from the region of Annaba

Annaba is a very industrialized region specialising in steel industries. The steel complex of Hadjar is located 12 kilometres south of the city of Annaba. The air pollution spreads over a radius of 6 km. According to [34], the complex annually releases into the atmosphere: 36890 tons of particles, 845 t of $NO_x$, 30895 t of $CO_x$, 2260 t of $SO_x$ and 3093 t of $NO_x$. The petrochemical station (ASMIDAL) produces fertilizers and pesticide products that have a big influence on air quality. 5 industrial zones, that contain hundreds of factories, are very close to the urban area and have a large impact on air pollution. The seaport is located in the centre of the city and attracts a lot of heavy transport, which also leads to deterioration in the air quality.

The local pollution agency network provided hourly data for a two-year period from 01/01/2003 to 31/12/2004. The concentrations of air pollutants that have been continuously monitored are: Ozone ($O_3$), Particulate Matter (PM10), Nitrogen Oxides ($NO_x$), and Sulphur Oxides ($SO_x$). The dataset also includes four meteorological parameters: Wind Speed (WS), Temperature (T) and relative Humidity (H). Daily rainfall measurements (RF) were also provided by the water management agency. The 2003 dataset was used for training the ANN and the 2004 dataset was used for validation; this helped us to assess the performance of the model. The pollutant concentration measurements are in microgram/$m^3$ and they have been normalised using equation (23).

$$V'_p = \frac{V_p}{(\max(V_p) - \min(V_p))}\quad (23)$$

Where $Vp$ is a parameter vector, $min$ and $max$ are functions that return the minimum and maximum values of the vector. Negative values, resulting from faulty measurements, were replaced using the mean of the previous and next values. It is impossible to discard faulty values since gaps in the time series will result in a data shift that affects the ANN training process leading to poor generalisation properties. Similarly, faulty (blank) measures for pollutants and weather parameters were replaced by an average of the $v$-$q$ and $u$+$q$ previous and future values respectively, with $u$ being the faulty sample and $q$ the number of values to take into consideration. This ensures the continuity and consistency of the time series and allows efficient training of the ANN predictors. Table 3 presents the statistical properties of the available data for different pollutants and weather parameters, for some parameters data are not available (N/A).

We defined a simulation scenario for the Annaba region using the parameters in table 4. The goal levels for pollutants concentration were fixed according to the air quality index goal. For this scenario we aimed to reach a very good air quality level (Goal air quality index=1). The initial values (at $t=0$) for pollutant concentration and climatic parameters were fixed according to the dataset.

| Parameter | 2003 mean | 2004 mean | 2003 STD | 2004 STD | Max value |
|---|---|---|---|---|---|
| PM10 µg/m³ | 51.70 | 27.76 | 51.66 | 26.38 | 508 |
| NOx µg/m³ | 14.50 | N/A | 25.01 | N/A | 435.0 |
| SOx µg/m³ | 7.60 | N/A | 14.78 | N/A | 190.0 |
| CO µg/m³ | 1.31 | N/A | 0.52 | N/A | 12.2 |
| O₃ µg/m³ | N/A | 42.27 | N/A | 64.58 | 688.0 |
| Wind Speed µg/m³ | 2.65 | 2.12 | 1.78 | 1.27 | 9.6 |
| Humidity (%) | 63.52 | 71.92 | 16.50 | 14.33 | 93.0 |
| Temperature (°C) | 18.96 | 16.82 | 7.76 | 6.30 | 42.1 |
| Rainfall (mm) | N/A | 2.96 | N/A | 9.27 | 73.9 |

Table 3: Statistical properties of the used dataset.

For the EG strategies we fixed the initial proportion of cooperating agents (agents choosing to decrease emissions) to $0.5$, this means that $50\%$ of the agents decrease their emission at $t=0$. The value of this parameter was chosen following the work of [14] and [15]. The proportion will change during the simulation according to the game outcome. The prediction was for the next 2 hours, the same as the simulation step. Each source emits according to its emission rate which cannot be higher than the maximum level defined in the simulation scenario. The position of sources was randomly generated and many sources are located in the same box.

| Parameter Name | Value |
|---|---|
| **Polluting activities and Policy parameters** | |
| Number of PM10 sources | 100 |
| Number of SO$_X$ sources | 100 |
| Number of NO$_X$ sources | 100 |
| Number of CO sources | 100 |
| Max emission rate | 2000 (gram/hour). |
| Goal PM10 level | 20 μ gram/m$^3$ |
| Goal SO$_X$ level | 30 μ gram/m$^3$ |
| Goal NO$_X$ level | 45 μ gram/m$^3$ |
| Goal O$_3$ level | 45 μ gram/m$^3$ |
| Number of memory steps ($L$) | 4 steps |
| Initial proportion of cooperating agents | 0.5 |
| **Environment parameters** | |
| Number of boxes | 20 |
| Temperature at t=0 | 12.7 (°C) |
| Humidity at t=0 | 71.0 % |
| Wind Speed t=0 | 2.4 m/s |
| PM10 at t=0 | 13.0 μ gram/m$^3$ |
| SO$_X$ at t=0 | 17.0 μ gram/m$^3$ |
| NO$_X$ at t=0 | 2.0 μ gram /m$^3$ |
| CO at t=0 | 0.5 μ gram /m$^3$ |
| O$_3$ at t=0 | 29.0 μgram /m$^3$ |
| Air Quality at t=0 | 2 ( Good) |
| Total simulation time | 4900 hours |
| $K$ Simulation step | 1 step = 2 hours |
| Prediction horizon | Next 2 hours |

Table 4: Parameter values of the simulation scenario.

## 4    Results and discussion:

We have built a simulator using the approach described above. We used the JADE agent framework [35] and ANN models from Encog [36]. We have defined 5 strategies: EG-CP (Evolutionary Game with Cumulative Penalty), EG-NCP (Evolutionary Game with No Cumulative Penalties), EG-NP (Evolutionary Game with No Penalty), CS (Centralized Strategy) and NC (No-Cooperation). The last one is used for comparison purposes. Using the parameters shown in table 4, we chose a strategy and ran the simulation 16 times. We then changed the strategy and ran the simulation again 16 times; since we have 5 strategies we obtain 80 simulations. The most explicative results are presented. For the CS and NC cases the simulator showed similar results for each run. For the EG strategies there were small differences between runs, especially concerning the proportion of cooperating agents. These changes are due to the random values used in the initialisation of some variables (neighbours rewards, first chosen action, weights, $k$ last actions and rewards). The comparison is done according to the air quality index. Results are expressed in terms of the number of occurrences of air quality index as illustrated in figure 2, for example the number of times the air quality index equals 1 (very good). Figure 3, shows the evolution of the air quality index over time. The CS gives the best performance. With the CS the air quality index moves rapidly from bad to average and then to good and finally stabilises at very-good (which is the goal fixed in the simulation scenario). The EG-CP moves the index from bad to average, when the equilibrium is reached it stabilises in good and never reaches a very-good index. The EG-NCP strategy moves the air quality from bad to average and never improves. When penalties are not used (EG-NP) the air quality stabilises at bad. When cooperation is not used (NC), agents act selfishly and do not care about the pollution, therefore, the air quality oscillates between bad and very-bad. As the agents reach their maximum emission rate we can observe an oscillation, which is caused by the climatic conditions. The only thing that affects the pollutant concentration is the climatic conditions (the emission rate is constant); these have a big influence and are captured with the ANN model.

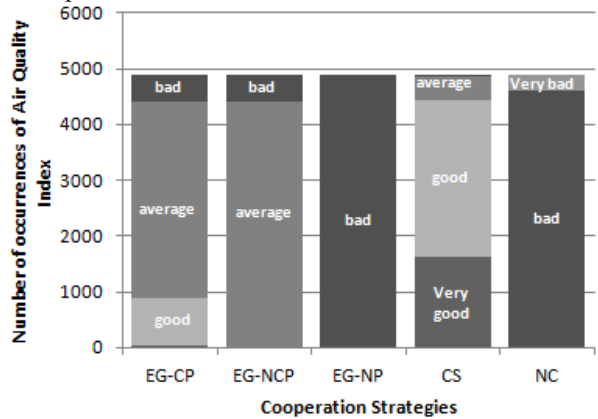

Figure 2: Air quality index using 5 different cooperation strategies

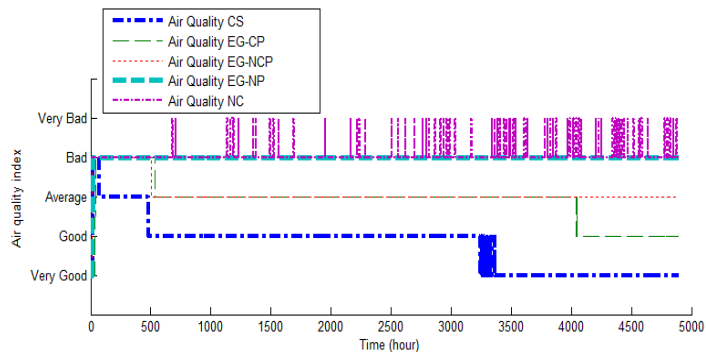

Figure 3: Air Quality index for 4900 hours.

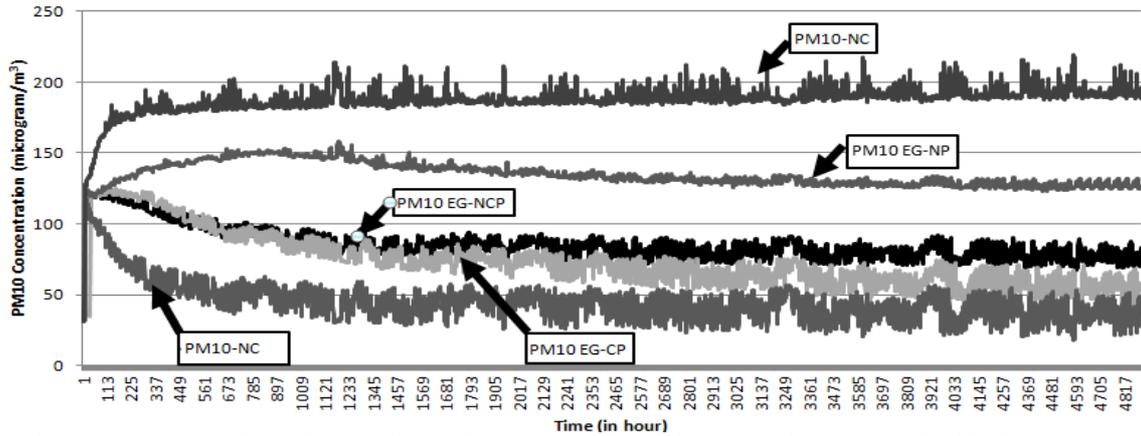

Figure 4: Concentration of PM10 for the four tested cooperation strategies compared with the no-cooperation strategy.

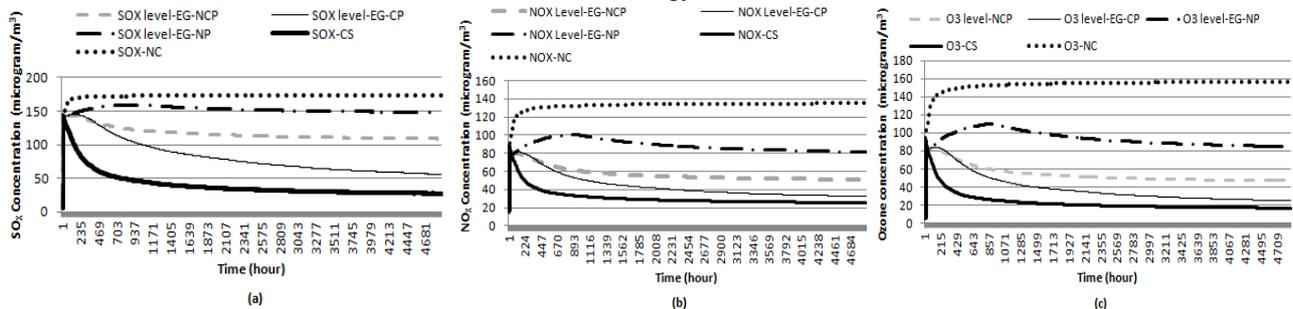

Figure 5 : Concentration of SO$_X$ (a), NO$_x$ (b) and O$_3$ (c) using the four strategies and the no-cooperation scenario.

Figure 4 shows the evolution of the PM10 concentration during the simulation time. The PM10 concentration shows many peaks compared with the other pollutants under the same climatic conditions. This is due to the dry nature of the weather in the Annaba area, with wildfires, and sandstorms coming from the great Sahara desert. These events have a big effect on the PM10 concentration but not on the other pollutants. The CS strategy takes less time to control the pollution level and keep it below the goal level defined in the simulation parameters. All of the EG strategies take longer, keeping it close to the goal level, but without ever reaching it. The penalising regulations have a big effect on the PM10 level. As illustrated, the EG-CP (cumulative penalising method) controls the pollution better than the non-cumulative one, and both methods perform better than the no-penalising strategy. The no-cooperation is presented in order to show the impact of cooperation on the PM10 level. Figure 5 shows the evolution of the SO$_x$, NO$_x$ and Ozone concentrations during the simulation time using four different cooperation strategies. The CS strategy gives the best performance since the pollution concentration rapidly decreases. The EG strategies show the same performance as for PM10 and the pollution level is widely influenced by the selected penalising method. The CP strategy appears to be the best one followed by the NCP. The pollution slowly decreases, but not enough to reach the goal level if penalisation is not used.

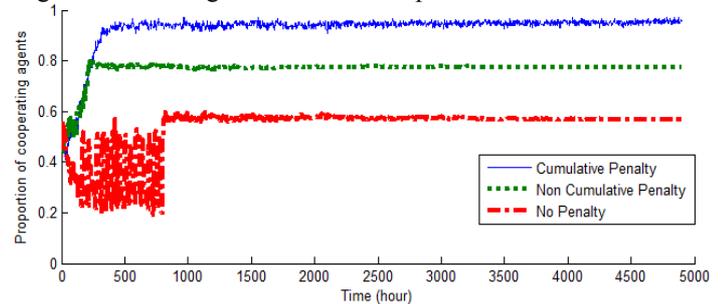

Figure 6: Proportions of cooperating agents for EG-CP, EG-NCP and EG-NP.

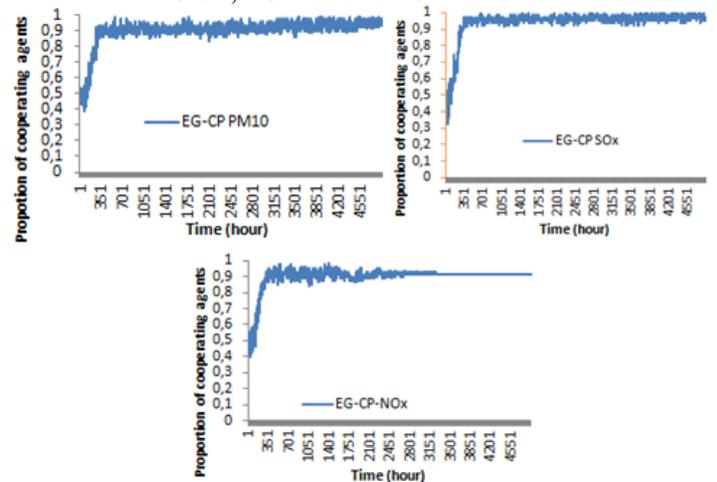

Figure 7: The proportion of cooperating agents according to the emitted pollutant for the EG-CP strategy.

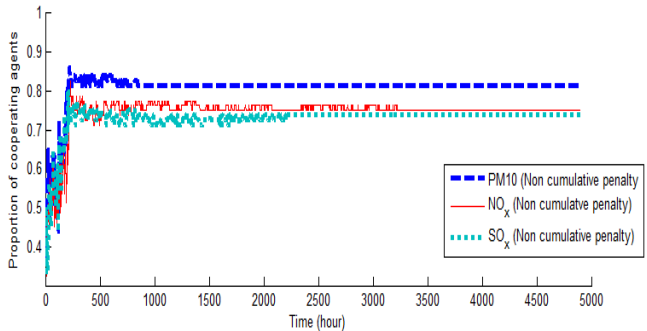

Figure 8, The proportion of cooperating agents for the EG-NCP strategy.

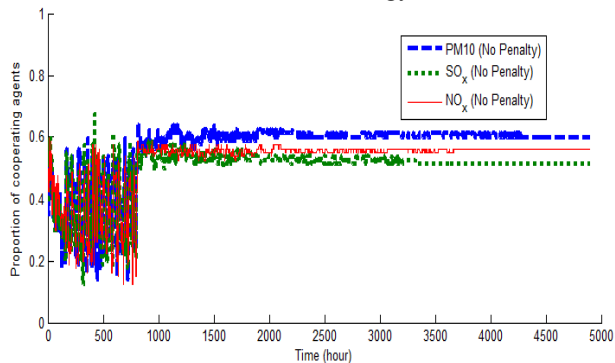

Figure 9: The proportions of cooperating agents, for the 4 groups of agents, when penalisation is not used.

The NC strategy gives the worst levels; when all agents are emitting pollution using their maximum emission rate, the pollutant level reaches alarming values and peak periods occur.

Figure 6 shows how the penalising method affects the proportion of cooperating agents. For the case of EG-CP the game equilibrium is reached at time step 387 and stabilises when the proportion of cooperating agents is between 0.93 and 0.95. The EG-NCP strategy stabilises early at time step 196 and oscillates between a cooperation ratio of 0.73 and 0.80, after which the equilibrium is fixed at 0.77. The EG-NP is the slowest; the equilibrium is reached at time step 808 with a cooperation ratio of 0.57. This happens because the agents are not penalised since the strategy does not include penalising methods. Figures 7, 8 and 9 show, respectively, the proportion of cooperating agents according to the pollutant for the three penalising strategies EG-CP, EG-NCP and EG-NP. The PM10 agents gives the highest cooperation ratio. This is because many peaks occur with this pollutant and the others pollutants cooperating ratio ($NO_x$ and $SO_x$) are influenced by the $O_3$ concentration. The more the pollutant exceeds the allowed level, the greater the proportion of co-operators. The equilibriums are disturbed by the pollution level, because, when the pollution level reaches a level above the goal level, agents start being penalised, and thus they tend to cooperate more.

# 5   Conclusions

Anthropogenic activities are among the main causes of pollution and environmental problems. These activities have to be included in the simulation models. Modelling the interaction between social and ecological components is a very important aspect. A MAS approach allows us to model the social network of human-beings sharing the exploitation of common environmental resources. Manipulating the behaviour at an individual and group level helps to gain more knowledge about the impact of human decision-making on pollution and makes the simulation more realistic. Studies treating air pollution are usually concerned with the physical aspects (concentration and dispersion of pollutant), and do not include human-decision factors on the emission sources.. In our approach, we model the decision-making activity of the air pollution emission source managers. This helps to investigate the conditions and regulations that may enhance and maintain the air quality.

We used a two stage air pollution modelling method: a GPD dispersion model and an ANN forecasting model. The ANN predictor uses climatic parameters and dispersal information provided by the GPD model to make predictions. This helped to introduce the effect of uncertainty caused by the weather and made the simulation more realistic. Five cooperation strategies were tested. The centralized cooperation strategy (CS) showed the best performance, surpassing the reward/penalty strategies. However, the CS strategy needs an effective communication network between emission sources controllers and the regulation agency. Also, we assume that emission controllers communicate exactly their emission rate, which is not always the case. The reward/penalty strategies seem to be more realistic; penalising the polluting agents according to their participation during peak periods has a big influence on their behaviours. As shown in the simulation results, it helps reducing the pollution level and affects the evolution of the pollutant. Thus, air pollution regulations have a big impact on pushing the emission source controllers to take their polluting activity seriously; this is especially important during the peak periods where climatic conditions cause the pollutants to stagnate.

To summarise our study helps to: (1) Model and introduce human decision-making concerning emission sources and the process of simulating air pollution evolution. (2) Evaluate the possible cooperation between the actors concerned in managing the air quality. (3) Have a prediction about the efficiency of regulation rules for preserving the air quality. (4) Investigate the impact on air quality of the decision to expand or establish a new emission points.

Our work aims to provide a decision-making tool to the air pollution control agencies that will help them evaluate the regulations and policies concerning air pollution control. The current version of the system deals only with point emission sources. In future versions we aim to include line and area sources. Line sources model the road activities, whereas area sources model the waste management and agricultural sources of pollution. If data becomes available in future it could be interesting to experiment with different time representations. Fortunately the multi-agent system approach allows us to easily change to a different scale of time representation in the same simulation. We can envisage using one time representation for decisions and another for monitoring. The first can help us to see long term impacts (e.g. investing in less polluting activities), and the second can help to see the short term changes.

The simulator may also be enhanced by including topographic aspects of regions since this has a big influence on the dispersion of air pollutants. In addition, including more agent personalities and exploring other cooperation strategies are also among our future plans. Our system is designed in a generic way and it could be adapted for other types of pollution such as water pollution. This could be done by changing the current dispersion and the prediction models to a water pollution dispersion model.